\documentclass[final,1p,times]{elsarticle}
\usepackage{hyperref}
\usepackage{amssymb}
\usepackage{amsmath}
\usepackage{color}
\usepackage{soul}
\usepackage{graphicx}
\begin{document}

\begin{frontmatter}

\title{Chukchi Myths perspective on Special Relativity}

\author[1,2]{Zurab K. Silagadze}

\affiliation[1]{organization={Novosibirsk State University},
            addressline={}, 
            city={Novosibirsk},
            postcode={630090},
            state={},
            country={Russia}
            }

\affiliation[2]{organization={Budker Institute of Nuclear Physics},
            addressline={}, 
            city={Novosibirsk},
            postcode={630090},
            country={Russia}
            }

\ead{silagadze@inp.nsk.su}

\begin{abstract}
The teaching of special relativity still follows Einstein's original two-postulate approach and thus recreates the relativistic revolution in the minds of students again and again, with all its attendant shocking and mysterious aspects. As Hermann Bondi long ago noted, such an approach, which emphasizes the revolutionary aspects of a theory rather than its continuity with earlier thought, ``is hardly conducive to easy teaching and good understanding". But what could be a better alternative? In 1923, the distinguished Russian ethnographer, linguist, and anthropologist Tan-Bogoraz described the striking similarities between the special theory of relativity and the mythology of Chukchi shamans. Inspired by this surprising observation, I assume that the basic concepts of relativity are not at all alien to our innate perception of time and space, and I propose an approach to the foundations of relativity that emphasizes absolute concepts such as proper time and causal cones rather than relative ones.
\end{abstract}

\end{frontmatter}

\section{Introduction}
Students experience difficulties when first introduced to the special theory of relativity(rele\-vant references on students' misconceptions about time in special relativity can be found in \cite{Alizzi_2022}), and some people, including Hermann Bondi \cite{Bondi_1966} and John Bell \cite{Bell_2001}, have proposed modernizing the teaching of special relativity to make the theory less paradoxical in the eyes of students.

Russian anthropologist Tan-Bogoraz in his work “Einstein and Religion: Application of the Principle of Relativity to the Study of Religious Phenomena” \cite{Bogoraz_1923} (see also \cite{Bogoraz_1925}), richly illustrated with amusing drawings by the Chukchi, provides the precise description of the clock paradox in a Chukchi legend:

``A certain shaman set out to wander distant lands, semi-mythical or even purely fairy-tale. After a year or two, after an indefinite period of time, he returns. He is still in full strength, in the prime of his health. But his home village has completely changed. His home has fallen into disrepair. He once left behind a wife and young son. But they have disappeared. He meets an old man with a gray beard on the road and asks him about his son. It turns out that this old man is his own son. Two years of travel through fairy-tale lands have flown by like an entire human life on earth. And the wanderer returns younger than his own son" \cite{Bogoraz_1923}.

It is remarkable how Bogoraz explains the similarity. ``What are the reasons for such a striking similarity between the constructs of modern physics and the traditions of prehistoric times? Clearly, primitive consciousness, like modern consciousness, utilizes the same way of perceiving the world, and both operate on the basis of identical relationships between subject and object, observer and observed. After all, modern mathematical physics, for all the ingenious complexity of its mathematical figures, ultimately simplifies and, as it were, returns to the semi-instinctive worldview of primitive consciousness, not yet purified and obscured by the formal scholasticism of nominal philosophy" \cite{Bogoraz_1923}.

In this essay, I will attempt to present the basic concepts of special relativity in a way that is consistent with the fundamental tenet of the shamanic myth: that absolute time does not exist, and that proper time, as perceived by humans and all objects, is non-integrable (in the sense that it depends on the trajectory through spacetime connecting two specific events in spacetime). In essence, this is a continuation and further development of the ideas of the work \cite{Alizzi_2022}.

\begin{figure}[htb]
    \centering
    \includegraphics[scale=0.8]{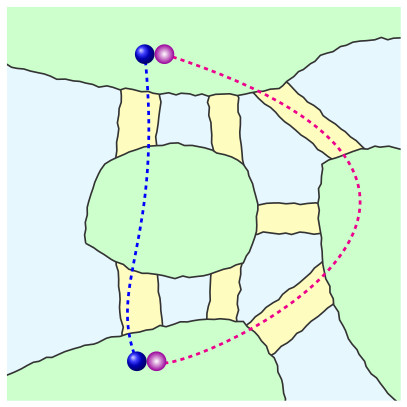}
     \caption{Hamster ball illustration of non-integrability of proper time. }
    \label{fig0}
\end{figure}

Perhaps the following artistic illustration will help to understand the non-integrability idea of the proper time, embedded in the Chukchi myth (see Fig.\ref{fig0}). Imagine that the internal clocks of the shaman and his son are represented by small hamster balls. Inside there are identical hamsters running at the same constant speed. As a result, the hamster balls roll along the surface of the ``ground” without slipping or twisting. Looking at the Fig.\ref{fig0}, it immediately becomes obvious that the number of revolutions made by the hamster balls depends not only on their initial and final positions, but also on the trajectory of their movement. Thus, such hamster balls allow to study the geometry of the ``land" (spacetime) \cite{Wise_2006}\footnote{in \cite{Wise_2006} the hamster ball is used to illustrate the idea of Cartan's geometry.}.

\section{Basic notions of special relativity}
In describing the basic concepts of the theory of relativity, I sacrifice some mathematical rigor and detail of the mathematical concepts used in favor of simplicity.

\subsection{Events and spacetime}
An event in special relativity is an abstraction analogous to a point in geometry or a material point in Newtonian mechanics, so it has neither spatial nor temporal extension. A collection of events constitute spacetime providing a domain of definition for variables describing particles and fields \cite{Ehlers_2006}. These basic concepts of relativity may perhaps appear less vague if we realize that similar concepts can be introduced into Newtonian physics. In a Newtonian universe, real, localized physical events occur at a specific place and time. So, to uniquely localize these events, we need four numbers $(t,x,y,z)$. The set of all such quadruples can be called four-dimensional space-time, the arena where all physical phenomena occur. The space-time formulations of Newtonian physics allow us to place the theories of Einstein and Newton on an equal footing, with virtually all the differences between them being due to differences in the structures of their symmetry groups – the Galilean group and the Lorentz group, respectively \cite{Havas_1964,Kuenzle_1972}. 

However, in the Newtonian case, the union of space and time is rather weak and illusory: space-time is a stack of three-dimensional hypersurfaces (spaces) threaded along the axis of absolute time. Therefore, we transfer this Newtonian construction of space-time to the special theory of relativity with reservations. First, we assume that space and time are no longer weakly bound, and replace space-time with spacetime, with the possible unification of space and time so tight that they may even cease to exist as separate entities \cite{Minkowski_1984}. Thus, although we still assume that spacetime is four-dimensional, we do not necessarily require that the four numbers needed to locate points in spacetime have spatiotemporal significance. Any coordinates that smoothly and one-to-one map some region of $\mathbb{R}^4$ onto the region of spacetime of interest will do. Understanding this fact is especially important in general relativity, whereas in special relativity it is customary to use coordinates with a clearly defined spatiotemporal  meaning (but even in special relativity nothing forces us to do this).

We typically partition spacetime using spatial and temporal coordinates and view the dynamics of particle interactions as unfolding in time, in a series of instants of the temporal coordinate. Dirac calls this form of dynamics the instantaneous form and points out that we are not obliged to choose it, since other forms of relativistic dynamics are also possible \cite{Dirac_1949}.

To summarize, events are points in spacetime, places where (idealized) real physical events can occur. Why do we call them "events" and not "points"? This is simply a historical tradition, perhaps originally connected to the desire to distinguish points in spacetime from points in ordinary space. Furthermore, the term "events" reminds us of the special role of time in the theory of relativity, and that the geometry underlying relativity is not ordinary geometry, but chronogeometry. From now on, we will use the terms "events" and "points" interchangeably.

To illustrate the ideas, all the drawings and considerations in this manuscript are made in a simplified $1+1$ dimensional spacetime.

\subsection{Causal structure and light cones}
Spacetime as a mere collection of events isn't very useful. We need to impose some additional structures on it to make it suitable for describing real physics. An example from geometry may elucidate this point \cite{Maudlin_2012}.

The most basic, fundamental level of geometric structure is the topological structure, which defines the meaning of continuity. The next level of geometric structure is the differentiable structure, which defines the meaning of smoothness and thus distinguishes smooth, continuous curves from curves with wedge-shaped fragments. Next we have projective structure  that distinguishes between straight lines and other lines. The affine structure introduces the concept of parallelism. Finally, the metric structure determines the distances between points and specifies the lengths of lines.

\begin{figure}[htb]
    \centering
    \includegraphics[scale=0.8]{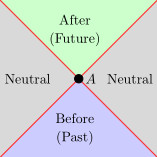}
     \caption{Absolute notions of ``Before", ``After" and ``Neutral".}
    \label{fig1}
\end{figure}

The most basic structure that we assume spacetime has is a causal structure. For a given event $A$, other events are divided into three categories. The first group is the set of events that $A$ can influence. These events constitute the future of event $A$, that is, the events that will occur after $A$. The past of event $A$ is defined similarly: the set of events that could have influenced $A$ and, therefore, occurred before $A$. The third group is the set of neutral events, not causally related to $A$. This division is shown schematically in Fig.\ref{fig1}.

Note that we do not yet have a concept of time: we have only defined such attributes of time as "Before" and "After", which partially order events in spacetime, introducing what is called a conical order in spacetime.

The first important difference between relativistic spacetime and Newtonian space-time is that the set of neutral with respect to $A$ events is four-dimensional. Consequently, we cannot assert that neutral events occur simultaneously with $A$: for any introduction of time coordinates in spacetime, if event $A$ is labeled by time $t$, then for the times of events neutral with respect to $A$ there will exist an entire interval $t_1<t<t_2$.

Events lying on the boundary between potentially causally related and causally unrelated (neutral) events with respect to $A$ form the so-called causal double cone with center at $A$. The set of all causal double cones imposes a conformal structure on spacetime (conformal because after introducing coordinates into spacetime, the double cones remain invariant under so-called conformal transformations).

The assumption of such a conformal structure of spacetime follows from our experience in describing physical systems with various interactions. Such systems are typically described mathematically by formulating an initial-value problem for a quasi-linear system of first-order hyperbolic partial differential equations, and any such system has its own causal cones, in the sense that any perturbation of the initial data propagates only within these cones. Einstein's equations in general relativity can also be represented as a symmetric hyperbolic system and, therefore, have causal cones \cite{Geroch_2010}. Since special relativity is a special case of general relativity, it inherits these causal cones. In fact, in modern physics, the causal cones of special relativity play a special role: other known physical systems either share their causal cones with special relativity, for example electromagnetism, or their causal cones lie inside the causal cones of special relativity. One can imagine the opposite situation - a democracy of causal cones, in which special relativity is just another physical theory, no different from our other theories described by symmetric hyperbolic systems, and its causal cones are not special at all \cite{Geroch_2010}. There may be many other systems, perhaps hidden sectors of the Universe, that use causal cones that lie outside the causal cones of special relativity, thereby supporting the propagation of superluminal signals \cite{Geroch_2010,Chashchina_2011,Chashchina_2021}. ``None of this would contradict our fundamental ideas about how physics is structured: an initial-value formulation, causal cones governing signals, etc." \cite{Geroch_2010}. However, at present nothing in our experimental data about nature requires such a revision.

The special role of light in special relativity is a kind of accident, a result of the Higgs mechanism, which leaves the photon massless. In any case, there is no evidence yet that the photon is not massless and, therefore, that the causal cones of electromagnetism do not coincide with the causal cones of special relativity. Consequently, from now on, we will refer to the causal cones of special relativity as light cones.

\subsection{Worldlines}
\begin{figure}[htb]
    \centering
    \includegraphics[scale=0.6]{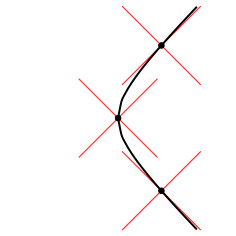}
     \caption{A timelike world line at any event on it lies inside the light cone of that event.}
    \label{fig2}
\end{figure}

A worldline is a smooth, continuous curve in spacetime. For a material object, a world line represents the set of all potential events that the object can experience. Such a world line is timelike: at any point, it lies within the light cone of that event (Fig.\ref{fig2}), since the light cone delimits the region over which a given material object can exert a causal influence. Worldlines of particles with zero mass (for example, a photon) lie on light cones. In the ordinary language of space and time, such particles transmit causal influence most quickly, and therefore their speed is the maximum speed in known nature. However, in order to speak meaningfully about the notion of speed and understand the meaning of the statement that the speed of light is the maximum possible speed of material objects, we must first introduce the concepts of space and time, that is, decompose spacetime into space and time. Nature does not provide an objective concept of speed, so starting the presentation of the special theory of relativity with Einstein's second postulate is not the best idea, as it is guaranteed to confuse students' minds. The four-dimensional presentation we advocate makes the conventional (stipulated in Einstein's own words) character of the second postulate obvious and makes it more accessible to students.
If spacetime is homogeneous, the light cones will be the same at all points. In the figures we will agree to depict them vertically. In the presence of gravity (when spacetime is curved), the light cones will be different at different points, but the statement about timelike worldlines, that they always lie inside the light cones at every point, will remain true.

\subsection{Ideal clocks and proper time}
On par with the concepts of spacetime and light cones, the concept of an ideal clock forms one more fundamental basis of the special theory of relativity. The basic assumption of this theory is that each timelike worldline of material objects with non-zero mass is associated with an ideal clock that measures the “length” along that worldline --- the proper time. Proper time plays the same role in special relativity as absolute time in Newtonian physics, and is therefore its successor. ``The ideal master clock of classical physics has shattered into a huge plurality of miniature ideal clocks in relativity, one for every timelike path" \cite{Callender_2017}.

All ideal clocks operate identically. They never "slow down" or "speed up," regardless of the mode of their motion (regardless of the timelike worldline they follow). They simply measure the proper time along the world line, and that is the definition of {\it an ideal clock}.  We can compare the ticking of two ideal clocks only when they are adjacent and stationary relative to each other. In all other cases, comparison is meaningless, since the theory of relativity has no global concept of "true" time. In some symmetric spacetimes (in particular, Minkowski spacetime), one can introduce a global time coordinate and compare the proper time of a particular clock with this coordinate time  --- this comparison is what is commonly called "time dilation" in special relativity.  However, nature does not care about our artificially constructed coordinate time and does not listen to the corresponding master clock when it comes to the dynamics of matter-energy \cite{Callender_2017}. The only objective time in the theory of relativity, the one to which nature heeds, is proper time.  However, unlike Newtonian physics, it is impossible to combine multitude of proper times into one master clock, and this is the most profound and radical change in our concept of time that special relativity brings about.

The proper time interval between two events in spacetime depends on the timelike worldline connecting these events (the so-called twin paradox). From a mathematical point of view, the infinitesimal proper time interval $d\tau$ is frame-independent but not integrable, whereas in Minkowski spacetime, for each congruence of inertial observers with parallel straight timelike worldlines (for each inertial frame of reference), one can introduce a coordinate time $t$ (see below), which imitates Newtonian time in the sense that the infinitesimal coordinate time intervals $dt$ are integrable (lead to path-independent line integrals), but they are not invariant with respect to frame changes.

\begin{figure}[htb]
    \centering
    \includegraphics[scale=0.6]{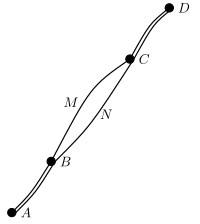}
     \caption{If ideal clocks separate at event $B$, follow different timelike worldlines $BMC$ and $BNC$, and reunite at event $C$, they will typically exhibit different elapsed proper times (the first clock effect). However, their tick rates will remain the same after reunion (absence of the second clock effect).}
    \label{fig3}
\end{figure}

From the four-dimensional perspective, differential aging of twins is a natural result of the non-integrable nature of proper time (see Fig.\ref{fig3}). If there is any paradox here, it is why this simple and basic effect of special relativity has generated such a vast and confusing literature. 
``The paradoxical nature of an age difference between the twins disappears once one stops thinking in terms of Newtonian pictures" \cite{Adler_1965}. However, numerous examples show that completely freeing oneself from the framework of Newtonian concepts is no easy task. For example, ``a good many physicists believe that this paradox can only be resolved by the general theory of relativity. They find great comfort in this, because they don't know any general relativity and feel that they don't have to worry about the problem until they decide to learn general relativity" \cite{Schild_1959}. In fact, the problem is not ignorance of general relativity, since among the quite a large number of physicists who defend the need for general relativity to resolve the twin paradox, there are such first-class experts on general relativity as Einstein himself \cite{Pesic_2003} and Pauli \cite{Pauli_1958}. The problem is that Newtonian concepts are deeply ingrained in the way we think about physics. Further evidence of this is the fact that Alfred Schild's excellent and lucid old paper \cite{Schild_1959}, which attempted to discuss the twin paradox in terms of non-Newtonian concepts, is little known.

In special relativity, when ideal clocks reunite, they tick identically (they measure the same elapsed proper time over the common interval $CD$ of their world lines in Fig.\ref{fig3}). The end of the shaman legend anticipates a more general situation: ``Recognizing his son, the father immediately falls on the spot and crumbles from decrepitude" \cite{Bogoraz_1923}. Thus, the analogy between the Chukchi myth and the theory of relativity fails here: according to the theory of relativity, he would have continued to age at the same rate as his son. Remarkably, in physics, such a more general theory, in which the rate of a clock depends on its history, was proposed in 1919 by Hermann Weyl in his unsuccessful attempt to unify electromagnetism and gravity. Experiments limit the existence of such so-called second-clock effect \cite{Lobo_2018} and render the classical version of Weyl's theory phenomenologically uninteresting, as Einstein had anticipated from the outset. However, when combined with quantum theory, Weyl's idea led to the birth of modern gauge theories, which form the basis of our description of fundamental interactions in the Standard Model \cite{Jackson_2001,Raifeartaigh_1998}.

Photons (and massless particles in general) do not carry ideal clocks, they have no sense of time: the interval between any two events that lie on the photon's world line is zero.

\subsection{Poincar\'{e}-Einstein simultaneity}
Unlike the absolute concepts of “Before” and “After”, the conical order does not define any unambiguous concept of simultaneity for separated events. Simultaneity must be defined by convention, and the following Poincar\'{e}-Einstein definition of simultaneity turns out to be the most convenient.

\begin{figure}[htb]
    \centering
    \includegraphics[scale=0.8]{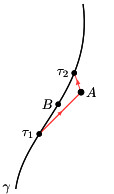}
     \caption{Illustration of the Poincar\'{e}-Einstein's definition of simultaneity.}
    \label{fig4}
\end{figure}

Let $\gamma$ be some worldline and an event $A$ is situated in the infinitesimal vicinity of $\gamma$ (Fig.\ref{fig4}). At proper time $\tau_1$ according to an ideal clock on $\gamma$ light signal is emitted which is immediatelly reflected at $A$ and received back on $\gamma$ at proper time $\tau_2$. The event $B$ on $\gamma$ is said to be simultaneous with $A$ if it corresponds to the proper time $\tau$ such that
\begin{equation}
    \tau=\frac{1}{2}(\tau_1+\tau_2)=\tau_1+\frac{1}{2}\left(\tau_2-\tau_1\right ).
    \label{eq1}
\end{equation}

In the language of space and time, Poincar\'{e}-Einstein simultaneity corresponds to the assertion that light travels in all directions at the same speed. Since this cannot be verified empirically (measuring the one-way speed of light requires synchronizing distant clocks, a concept of simultaneity that nature does not provide), other conventions of simultaneity are possible. In particular, instead of (\ref{eq1}), one can take \cite{Reichenbach_1958,Grunbaum_1973}
\begin{equation}
    \tau=\tau_1+\epsilon\left(\tau_2-\tau_1\right ),\;\;0<\epsilon<1.
    \label{eq2}
\end{equation}
Sometimes the limiting cases $\epsilon=0$ and $\epsilon=1$ are excluded because they correspond to an infinite one-way speed of light, but their preservation corresponds to the interesting possibility of frame-independent (not relative) simultaneity in special relativity \cite{Leubner_1992} (apparently, the possibility that frame-independent simultaneity is consistent with special relativity was first recognized by Frank Robert Tangherlini in his 1958 PhD dissertation \cite{Chashchina_2016}.

The widespread belief that the relativity of simultaneity is an integral feature of special relativity is false. Poincar\'{e}-Einstein simultaneity is defined relative to the worldline $\gamma$ and depends on it:  if we choose a different worldline $\gamma^\prime$ passing through event $B$, events $A$ and $B$ will no longer be simultaneous relative to this new worldline. In this respect, simultaneity is, of course, relative. However, the truly innovative and revolutionary aspect of relativity is the recognition of the absence of objective simultaneity in nature. The fact that it becomes relative in the definition of Poincar\'{e}-Einstein simultaneity is of secondary importance. After all, nothing but the simplicity of the subsequent mathematical description forces us to accept the standard Poincar\'{e}-Einstein simultaneity $\epsilon=1/2$ \cite{Minguzzi_2002}. But mathematical simplicity is a great thing, and we are relieved to accept the Poincar\'{e}-Einstein convention in the remainder of the manuscript.

\subsection{Radar coordinates}
\begin{figure}[htb]
    \centering
    \includegraphics[scale=0.8]{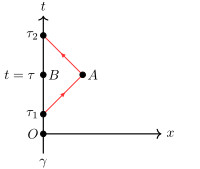}
     \caption{Radar time coordinates are defined with respect to the worldline $\gamma$.}
    \label{fig5}
\end{figure}

The simultaneity convention allows us to introduce spatiotemporal coordinates, at least in the immediate vicinity of some event $O$. Let the world line $\gamma$ of some observer pass through the point $O$. Since we are currently considering only an infinitesimal neighborhood of the point $O$, we can assume that $\gamma$ is a line segment. Let the zero of the proper time on $\gamma$ correspond to the point $O$.  Along $\gamma$ the time coordinate $t$ coincides with the proper time $\tau$. For an event $A$ near the world line $\gamma$, the proper time $\tau$ of the event $B$ on $\gamma$, which is Poincar\'{e}-Einstein simultaneous with $A$, is chosen as the time coordinate $t$ (Fig.\ref{fig5}).

The $x$-coordinate of $A$ is taken to be the distance between simultaneous events $A$ and $B$, defined as $x=\frac{1}{2}(\tau_2-\tau_1)$. This definition assumes that distance has the dimension of time. If we want distance to have the usual dimension, we must explicitly introduce the speed of light $c$ into the definition of distance: $x=\frac{c}{2}(\tau_2-\tau_1)$. However, $c=1$ is a convenient choice, since we can always restore $c$ in the formulas, if desired, based on dimensional considerations.

Therefore, the event $A$ with respect to the observer $\gamma$ has the coordinates
\begin{equation}
    t=\frac{1}{2}(\tau_1+\tau_2),\;\;\;x=\frac{1}{2}(\tau_2-\tau_1).
    \label{eq3}
\end{equation}
As a result, the observer $\gamma$ divides spacetime into space and time. In general, the described procedure can only be performed if $A$ is close to $\gamma$.  However, if spacetime is flat (not curved) and $\gamma$ is a straight line, then such a construction with light signals can, in principle, be performed for all events, thereby introducing coordinates throughout the whole spacetime.

The special theory of relativity in the traditional sense studies flat spacetime (Minkowski spacetime) from the point of view of inertial observers with straight world lines. In fact, the restriction to inertial observers is not strictly necessary, since real observers are generally non-inertial, but such an extension of traditional special relativity, while remaining a study of flat Minkowski spacetime, requires a more complex mathematical apparatus \cite{Gourgoulhon_2013}.

\subsection{Public time and public space. Inertial coordinates (frames)}
\begin{figure}[htb]
    \centering
    \includegraphics[scale=0.6]{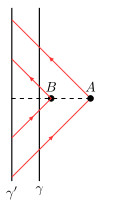}
     \caption{The parallel world lines $\gamma$ and $\gamma^\prime$ share the same concept of simultaneity.}
    \label{fig6}
\end{figure}

The space and time that the observer (particle) $\gamma$ got when dividing space-time into space and time can be called the private space and private time of the observer $\gamma$, since not all observers will agree with such a division (for a given observer, the natural direction of the time coordinate is along his/her world line). However, observers whose worldlines are parallel to $\gamma$ will agree with this division of spacetime, since they share with $\gamma$ the same notion of simultaneity: if events $A$ and $B$ have the same time coordinate on $\gamma$, they will have the same time coordinate on $\gamma^\prime$ if $\gamma^\prime\parallel\gamma$ (Fig.\ref{fig6}).
As a result, such observers can synchronize their clocks, and the private time and private space of the observer $\gamma$ are elevated to public time and public space for all such observers (we use Milne's terminology \cite{Milne_1948} when talking about the concepts of private and public time and space). In special relativity, the congruence (collection) of such parallel worldlines (observers) filling all of Minkowski spacetime forms what is called an inertial frame of reference, and their common (public) radar coordinates are the coordinates of this inertial frame of reference.

Radar coordinates in general spacetime are known as M\"{a}rzke-Wheeler coordinates \cite{Marzke_1964}. However, in the context of special relativity they were originally introduced by Alfred Robb \cite{Robb_1911}.

When we talk about private time for a given inertial observer, we mean the global time coordinate that this observer introduced by dividing spacetime into space and time, and not the proper time of this observer. By public time we mean a global time coordinate that corresponds to the division of space-time into space and time shared by a special class of privileged observers (for example, those with parallel worldlines). Private and public times can be drastically different, as illustrated by the Milne model (for details, see) \cite{Alizzi_2022}.

\subsection{Interval}
\begin{figure}[htb]
    \centering
    \includegraphics[scale=0.8]{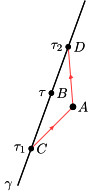}
    \includegraphics[scale=0.8]{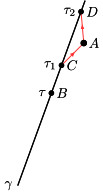}
    \includegraphics[scale=0.8]{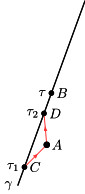}
     \caption{Robb-Geroch construction for determining the relativistic interval. If $A$ and $B$ cannot be causally related (left figure), $\tau_2-\tau$ and $\tau_1-\tau$ have different signs. Otherwise (right figures), $\tau_2-\tau$ and $\tau_1-\tau$ have the same signs.}
    \label{fig7}
\end{figure}

Let $A$ and $B$ be nearby events in spacetime, and $\gamma$ be a worldline passing through $B$.
At the proper time instant $\tau_1$, a light signal is emitted (event $C$), which is reflected from $A$ and received again on $\gamma$ at the proper time instant $\tau_2$ (event $D$). Since $A$ and $B$ are assumed to be very close, all this light signaling occurs in their immediate vicinity, and $\gamma$ can be considered a line segment in this region (Fig.\ref{fig7}).
If $B$ is simultaneous with $A$ with respect to $\gamma$, then $\tau=\frac{1}{2}(\tau_1+\tau_2)$, and the distance between $A$ and $B$ in this case is $l=\frac{1}{2}(\tau_2-\tau_1)=\tau_2-\tau=\tau-\tau_1$. Therefore, $l^2=(\tau_2-\tau)(\tau-\tau_1)$. The last relation makes sense even if $\tau\ne\frac{1}{2}(\tau_1+\tau_2)$. In particular, its sign may serve as an indicator of the causal relationship between $A$ and $B$ (see Fig.\ref{fig7}) \cite{Geroch_1972,Geroch_1981}, hinting that this quantity may have a deep meaning in the theory of relativity.  This is not surprising, since in fact $s_{AB}^2=(\tau_2-\tau)(\tau-\tau_1)$ turns out to be the (relativistic) interval between events $A$ and $B$, well known in the canonical presentation of special relativity. The advantage of this definition of the interval is that it becomes a direct generalization of the concept of Euclidean distance to spacetime, and, as a rule, such a definition is used in the literature on general relativity. However, if we want proper time to measure the "length" of a timelike worldline, it is better to define $s^2_{AB}$ with the opposite sign:
\begin{equation}
 s_{AB}^2=(\tau_2-\tau)(\tau_1-\tau).
 \label{eq4}
\end{equation}
Indeed, suppose that $A$ tends to $\gamma$. Then $\tau_1$ tends to $\tau_2$, and by (\ref{eq4}), in the limiting case when $B$ is on $\gamma$, $s_{AB}=|\tau_2-\tau|$ is simply the proper time interval between the corresponding events on $\gamma$, measured by an ideal clock.

This definition of the relativistic interval is originally due to Robb \cite{Robb_1911,Robb_1936} and is unfortunately not as well known as it deserves. Apparently, this form of the relativistic interval has been rediscovered several times \cite{Alizzi_2022} (see also \cite{Marzke_1964}). We call it the Robb-Geroch interval to emphasize Geroch's important contribution to elucidating the structure of spacetime in relativity theory \cite{Geroch_1972,Geroch_1981}. Interestingly, the Robb-Geroch interval admits a simple Finslerian generalization \cite{Sagaydak_2021}.

\subsection{Interval in terms of inertial coordinates}
Unlike the concept of simultaneity, which depends on the world line $\gamma$ used in its definition and is therefore a relative, conventional notion, an interval must not depend on the world line $\gamma$ used in its definition if we want it to represent a "distance" in spacetime. The interval $s_{AB}$ can depend only on the coordinates (inertial or not) of events $A$ and $B$ --- it must be some single-valued function of these coordinates. Let's see how this is possible.

\begin{figure}[htb]
    \centering
    \includegraphics[scale=1.0]{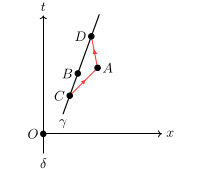}
     \caption{Robb-Geroch diagram for calculating the $s_{AB}$ interval in terms of radar coordinates $A$ and $B$.}
    \label{fig8}
\end{figure}

Suppose that an observer $\delta$ with a vertical worldline entered radar coordinates with the origin at event $O$, shared them via clock synchronization with other observers, also with vertical worldlines, and thereby made these coordinates public, forming coordinates of an inertial reference frame. To find the interval $s_{AB}$, we use the already familiar Robb-Geroch construction,  based on the worldline $\gamma$ passing through event $B$ (Fig.\ref{fig8}).

Let $A$ and $B$ be two infinitely close events with coordinates $(t,x)$ and $(t+\Delta t,x+\Delta x)$, respectively, and let the world line $\gamma $ passing through $B$ be given by the equation $x=x_0+\beta t$.  A light signal is emitted from point $C$ with coordinates $(t_1,x_1)$, which is reflected back at $A$ and again hits the worldline $\gamma$ at point $D$ with coordinates $(t_2,x_2)$. Since in radar coordinates (\ref{eq3}) the speed of light is equal to unity, then for the segment $CA$ of the null-worldline along which the light propagates, we can write $t+\Delta t-t_1=x+\Delta x-x1$. But $CB$ is part of the world line $\gamma$, and therefore $x-x_1=\beta(t-t_1)$. From these two equations it follows that
\begin{equation}
(1-\beta)(t-t_1)=\Delta x-\Delta t.
\label{eq5}
\end{equation}
Similarly, since $AD$ is another null-worldline, $t_2-t-\Delta t=x+\Delta x-x_2$. But $BD$ also lies on $\gamma$, and therefore $x_2-x=\beta(t_2-t)$. Then it follows that
\begin{equation}
(1+\beta)(t_2-t)=\Delta x+\Delta t.
\label{eq6}
\end{equation}
From (\ref{eq5}) and (\ref{eq6}) we get
\begin{equation}
(t_2-t)(t_1-t)=\frac{(\Delta t)^2-(\Delta x)^2}{1-\beta^2}.
\label{eq7}
\end{equation}
If $\Delta t$ is an infinitesimal coordinate interval, the corresponding proper time interval along $\gamma$ will be some function of $\Delta t$: $\Delta \tau=f(\Delta t)\approx f(0)+f^\prime(0)\Delta t$. But $\Delta \tau=0$ if $\Delta t=0$ (for example, if points $C$ and $B$ coincide, then both $t-t_1=0$ and $\tau-\tau_1=0$). Therefore, $f(0)=0$. $f^\prime(0)$ may depend on the worldline $\gamma$. Since our entire construction takes place in an infinitely small neighborhood of $A$ and $B$, we can assume that all worldlines passing through $B$ are straight line segments differing only in one parameter $\beta$, and write $f^\prime(0)=k_\beta$, that is, $\Delta \tau=k_\beta\Delta t$. Then $t_2-t=(\tau_2-\tau)/k_\beta$, $t_1-t=(\tau_1-\tau)/k_\beta$, and since $s_{AB}^2=(\tau_2-\tau)(\tau_1-\tau)$, from (\ref{eq7}) we obtain
\begin{equation}
s^2_{AB}=\frac{k^2_\beta}{1-\beta^2}\left [(\Delta t)^2-(\Delta x)^2\right ].
\label{eq8}
\end{equation}
The interval $s_{AB}$ will not depend on the worldline $\gamma$ (on the parameter $\beta$) if
\begin{equation}
\frac{k^2_\beta}{1-\beta^2}=\mathrm{const}.
\label{eq9}
\end{equation}
When $\beta=0$, the worldline $\gamma$ becomes parallel to the observer's worldline $\delta$, and as a consequence, shares a common public time $t$ with $\delta$. Thus, in this case $\Delta\tau=\Delta t$, and hence $k_0=1$. Therefore, taking $\beta=0$ in (\ref{eq9}), we see that $\mathrm{const}=1$ and $k_\beta=\sqrt{1-\beta^2}$.

As we see, from (\ref{eq8}) follows the familiar expression $ds^2=dt^2-dx^2$ of the Minkowski metric in inertial coordinates, but this requires the ``time dilation" relation between the coordinate and proper time intervals:
\begin{equation}
d\tau=\sqrt{1-\beta^2}\,dt=\frac{dt}{\gamma},\;\;\;\gamma=\frac{1}{\sqrt{1-\beta^2}}.
\label{eq10}
\end{equation}
It is this connection between proper and coordinate time intervals that reflects the true meaning of the notorious phrase ``a moving clock slows down". Both "time dilation" and "moving clocks slow down" are unfortunate historical artifacts, as they invariably confuse students. We recommend avoiding their use when teaching the theory of relativity.

Another interesting and useful observation from Fig.\ref{fig8} is that the Euclidean length of $CB$ is 
\begin{equation}
\hspace*{-10mm}
|CB|=\sqrt{(t-t_1)^2+\beta^2(t-t_1)^2}=\sqrt{1+\beta^2}(t-t_1)=\sqrt{\frac{1+\beta^2}{1-\beta^2}}\,(\tau-\tau_1). 
\label{CB_length}
\end{equation}
Therefore, to transform the Euclidean length of a segment of some timelike worldline into the corresponding proper time interval, we must divide this Euclidean length by a "calibration factor" \cite{Inverno_1992}
\begin{equation}
k=\sqrt{\frac{1+\beta^2}{1-\beta^2}}.
\label{eq11}
\end{equation}
To summarize this chapter, we can say that the Robb-Geroch interval is an ordinary relativistic interval in coordinate-free form.
 
\subsection{Primed axes and Minkowski diagram}
Suppose an observer with a world line $\delta$ has entered radar coordinates, and another observer has a world line $\delta^\prime$, which in these coordinates is given by the equation $x=\beta t$. The intersection event $O$ can serve as a common origin for their proper times, and for the moving observer, the new temporal axis $t^\prime$ is directed along his world line $\delta^\prime$ (Fig. 10).
\begin{figure}[htb]
    \centering
    \includegraphics[scale=1.0]{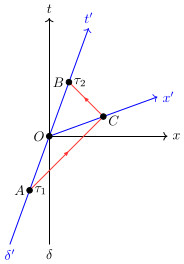}
    \caption{Event $C$ on the $x^\prime$ axis is simultaneous with event $O$ at the origin. }
    \label{fig10}
\end{figure}

But what is the direction of the new spatial axis $x^\prime$? The $x^\prime$ axis represents the set of events that occur simultaneously relative to the worldline $\delta^\prime$ with the event $O$ at the origin. Let us assume that event $C$ with coordinates $(t,x)$ lies on the $x^\prime$ axis. Since $\tau_O=0$, the simultaneity of $O$ and $C$ implies $\tau_1+\tau_2=0$, that is, $\tau_1=-\tau_2$ and, consequently, $t_1=-t_2$. Therefore, point $A$, from which the light signal was emitted in the direction of $C$, has coordinates $(-t_2,-\beta t_2)$, and point $B$, where the signal was received back on $\delta^\prime$, has coordinates $(t_2,\beta t_2)$. $AC$ is the worldline of the photon. Therefore, $t-(-t_2)=x-(-\beta t_2)$, and consequently, $x-t=(1-\beta)t_2$. But $CB$ is also the worldline of the photon. Therefore, $t_2-t=x-\beta t_2$, or $x+t=(1+\beta)t_2$.  As we see, $(x+t)/(x-t)=(1+\beta)/(1-\beta)$. It follows that the $x^\prime$ axis is defined by the equation $t=\beta x$ and makes the same angle $\arctan(\beta)$ with the $x$ axis as the $t^\prime$ axis makes with the $t$ axis.

In the Minkowski diagram, where both $(t,x)$ and $(t^\prime,x^\prime)$ event coordinates can be represented, we can trust our Euclidean intuition about projective and affine notions. This is so because both the Lorentz transformations (discussed in the next subsection), which leave the Minkowski line element $ds^2=dt^2-dx^2$ invariant, and the rotations, which leave the Euclidean line element $dl^2=dt^2+dx^2$ invariant, are linear transformations, and linear transformations take a point to a point, a straight line to a straight line, parallel lines to parallel lines, and preserve incidence \cite{Schild_1959}. However, when it comes to metric concepts, the difference in signs between Minkowski and Euclidean line elements becomes important, and our Euclidean intuition can no longer be trusted. For example, the $t^\prime$ and $x^\prime$ axes in Fig.\ref{fig10} are mutually orthogonal in Minkowski geometry, but clearly not in Euclidean geometry. Equal lengths in Minkowski geometry will not appear equal on a Minkowski diagram (for example, any segment of a photon's worldline has zero Minkowski length). However, the ``calibration factor" (\ref{eq11}) comes to the aid of our Euclidean intuition, since it can be used to transform the Euclidean length of any segment of a timelike worldline into the corresponding proper time.

The ``calibration factor"  is a tool that helps one correctly ``read" a Minkowski diagram, which is a symbolic representation of Minkowski spacetime on a Euclidean sheet of paper. Typically, this factor is obtained either using Lorentz transformations \cite{Morin_2017} or invariant calibrating hyperbolas \cite{Rindler_2006}.

\subsection{Lorentz transformations}
\begin{figure}[htb]
    \centering
    \includegraphics[scale=1.2]{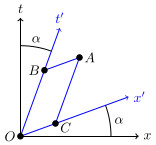}
     \caption{Obtaining Lorentz transformations for the event $A$.}
    \label{fig11}
\end{figure}

Recall that in special relativity, an inertial frame of reference is a congruence of parallel timelike worldlines, and all observers in this congruence share the same public temporal and spatial coordinates. Let some event $A$ in an inertial frame of reference $K$ have coordinates $(t,x)$. What coordinates will it have in another inertial frame of reference $K^\prime$, whose worldlines are inclined in $K$ by an angle $\alpha=\arctan(\beta)$, where $\beta$ is the relative velocity of $K$ and $K^\prime$? To find the coordinates of $A$ in $K^\prime$, it is necessary to draw lines on the Minkowski diagram starting at $A$ and parallel to the primed coordinate axes $t^\prime$ and $x^\prime$ until they intersect with these axes.  Then the coordinate $t^\prime$ of event $A$ will be equal to the proper time of the intersection point $B$, and $x^\prime$ will be the distance between simultaneous events $C$ and $O$ (see Fig.\ref{fig11}).

If event $B$ in the system $K$ has coordinates $(t_1,\beta t_1)$, then since $BA$ is inclined by the angle $\alpha$ with respect to the axis $x$, we will have $\beta=\tan{\alpha}=(t-t_1)/(x-\beta t_1)$, or $(1-\beta^2)t_1=t-\beta x$. Euclidean length of $OB$ is
\begin{equation}
|OB|=\sqrt{t_1^2+\beta^2 t_1^2}=\sqrt{1+\beta^2}\, t_1=\frac{\sqrt{1+\beta^2}}{1-\beta^2}\left (t-\beta x\right ).
\label{OB_length}
\end{equation}
This Euclidean length must be divided by the calibration factor $k(\beta)$ from (\ref{eq11}) to get the corresponding proper time
\begin{equation}
t^\prime=\frac{|OB|}{k}=\frac{1}{\sqrt{1-\beta^2}} \left (t-\beta x\right )=\gamma \left (t-\beta x\right ).
\label{LT1}
\end{equation}
Similarly, if point $C$ has coordinates $(\beta x_2,x_2)$ in $K$, then $\beta=\tan{\alpha}=(x-x_2)/(t-\beta x_2)$, since $CA$ is inclined by an angle $\alpha$ relative to the $t$-axis. The Euclidean length of $OC$ is $|OC|=\sqrt{1+\beta^2}\,x_2$. To transform this Euclidean length into the distance between simultaneous (in $K^\prime$) events $C$ and $O$, we must divide it by the same calibration factor (\ref{eq11}), since the distance between simultaneous events is defined through the proper time (see \ref{eq3}). As a result we get
\begin{equation}
x^\prime=\frac{|OC|}{k}=\frac{1}{\sqrt{1-\beta^2}} \left (x-\beta t\right )=\gamma \left (x-\beta t\right ).
\label{LT2}
\end{equation}
In (\ref{LT1}) and (\ref{LT2}) we recognize the usual form of the Lorentz transformations (with $c=1$). How to derive the transformation law for transverse coordinates in the same manner is described in \cite{Alizzi_2022}.

\subsection{Twin paradox}
\begin{figure}[htb]
    \centering
    \includegraphics[scale=0.7]{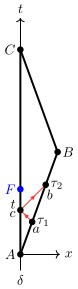}
    \hspace*{10mm}
    \includegraphics[scale=0.7]{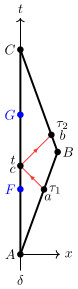}
    \hspace*{10mm}
    \includegraphics[scale=0.7]{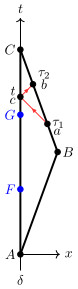}
     \caption{Robb-Geroch construction for determining the radar coordinates of rocket-bound twin.}
    \label{fig9}
\end{figure}

The literature on the twin paradox is vast (see, for example, \cite{Marder_1971},\cite{Mcdonald_2020}), probably because ``it is generally easier to write a paper on the 'clock paradox' than to understand one written by someone else" \cite{Mcdonald_2020}. Everyone agrees that the Earth-bound twin will find that his rocket-riding twin ages more slowly as time goes on. The confusion begins when people try to understand what the situation looks like from the perspective of the twin flying on the rocket. The conceptual framework advocated in this article provides a very simple answer to the question of how the twin riding the rocket calculates the other twin's proper time using radar coordinates associated with the rocket's world line. Our approach was influenced by \cite{Schild_1959} and is conceptually close to the approaches of \cite{Pauri_2000} and \cite{Dolby_2001}, although we hope it is much simpler. We'll consider the simplest, idealized case of the paradox depicted in Fig.\ref{fig9}, where $AC$ is the worldline of the Earth-bound twin, and $ABC$ is the worldline of the twin riding on a rocket. Since we want to demonstrate only the most important points, we won't worry about the infinite accelerations at certain points on the worldline of the twin riding on a rocket —-- we can always smooth out these sections of the worldline, which will complicate the calculations somewhat but won't change their essence.

Suppose that the twin in the rocket wants to assign his/her radar time coordinate to an event $c$ on the Earth-bound twin's worldline, which has coordinates $(t,0)$ on the coordinate map associated with the worldline $\delta$ coinciding with the Earth-bound twin's worldline. To do this, he/she must find an event on his/her worldline that is simultaneous with $c$. Thus, he/she emits a light signal at point $a$ and, after immediate reflection from $c$, receives it back at point $b$ (left figure on Fig.\ref{fig9}). Since $ac$ is the photon's world line, we can write $t-t_a=\beta t_a$, where $(t_a,\beta t_a)$ are the coordinates of $a$ and $\beta$ is the velocity of the moving twin (so the world line $AB$ is given by the equation $x=\beta t)$. Similarly, considering another null-worldline $cb$, we can write $t_b-t=\beta t_b$. Therefore, $t_a=t/(1+\beta)$ and $t_b=t/(1-\beta)$. To transform the Euclidean lengths $|Aa|=\sqrt{1+\beta^2}\,t_a$ and $|Ab|=\sqrt{1+\beta^2}\,t_b$ into the corresponding proper times $\tau_1$ and $\tau_2$, we must divide them by the calibration factor $k$ from (\ref{eq11}). As a result, we obtain $\tau_1=t_a/\gamma$ and $\tau_2=t_b/\gamma$. Consequently, the twin on the rocket assigns to the event $c$ the time coordinate
\begin{equation}
 t^\prime=\frac{1}{2}(\tau_1+\tau_2)=\frac{t}{2\gamma}\left (\frac{1}{1+\beta}+\frac{1}{1-\beta}\right )=\gamma t,\;\;\;t\le (1-\beta)T.
 \label{eq12}
\end{equation}
So far so good, we got what we expected: the twin on the rocket can think that the twin on Earth is aging $\gamma$ times slower. However, in this way the coordinate $t^\prime$ can be assigned to events on $AC$ only up to the point $F$, for which $b$ coincides with $B$, at which the rocket abruptly changes the direction of its velocity to the opposite. If the entire journey lasted $2T$ of the Earth-bound twin's proper time, so that point $B$ has coordinates $(T,\beta T)$ based on the $\delta$-worldline, the condition $t_b=T$ will yield $t=(1-\beta)T$, which will be the time coordinate of point $F$ on the coordinate map associated with the $\delta$-worldline.

After point $F$ the reflected signal will be received on the world line $BC$, which is given by the equation $x=\beta(2T-t)$ in unprimed coordinates (the middle figure on Fig.\ref{fig9}). Therefore, since $(t_b,\beta(2T-t_b))$ are the coordinates of event $b$, and $cb$ is the world line of the photon, we can write $t_b-t=\beta(2T-t_b)$. This gives $t_b=(t+2\beta T)/(1+\beta)$, while $t_a$ and $\tau_1$ are given by the previous expressions $t_a=t/(1+\beta)$ and $\tau_1=t_a/\gamma$. Since $|AB|=\sqrt{1+\beta^2}\,T$ and $|Bb|=\sqrt{(t_b-T)^2+[\beta T-\beta(2T-t_b)]^2}=\sqrt{1+\beta^2}(t_b-T)$, for $\tau_2$ we get $\tau_2=(|AB|+|Bb|)/k=t_b/\gamma$. Therefore, in this case the rocket riding twin assigns to the event $c$ the time coordinate
\begin{equation}
\hspace*{-10mm}
t^\prime=\frac{1}{2}(\tau_1+\tau_2)=\frac{1}{2\gamma}(t_a+t_b)=\frac{t+\beta T}{\gamma(1+\beta)},
\;\;\;(1-\beta)T\le t \le (1+\beta)T.
\label{eq13}
\end{equation}   
Such assignments continues until the event $G$ is reached for which event $a$ coincides with the event $B$. The condition $t_a=T$ gives the time coordinate of the event $G$ as $t=(1+\beta)T$. After event $G$ is passed, both events $a$ and $b$ lie on worldline $BC$. We again use the fact that $ac$ is the world line of the photon and write $t-t_a=\beta(2T-t_a)$. Therefore, $t_a=(t-2\beta T)/(1-\beta)$, while as before $t_b=(t+2\beta T)/(1+\beta)$. Knowing the coordinates of points $B$, $a$ and $b$, we can find the Euclidean lengths $|Ba|=\sqrt{1+\beta^2}(t_a-T)$ and $|Bb|=\sqrt{1+\beta^2}(t_b-T)$ and use them to find the required proper time:
\begin{equation}
\hspace*{-10mm}
t^\prime=\frac{1}{2}(\tau_1+\tau_2)=\frac{1}{2k}(2|AB|+|Ba|+|Bb|)=\frac{1}{2\gamma}(t_a+t_b)=\gamma t-2\beta^2\gamma\, T,  
\label{eq14}
\end{equation} 
when $(1+\beta)T\le t\le 2T$. 
Equations (\ref{eq12}), (\ref{eq13}) and (\ref{eq14}) completely cover the world line $AC$ of the Earth-bound twin and assign a time coordinate $t^\prime$ to each event on it. If we were to follow the old tradition (which, unfortunately, still abounds), we would say that, compared to the time of the twin riding the rocket, the clock of the twin on Earth slows down by the standard $\gamma$ factor as he/she traverses the $AF$ segment of the $AC$ worldline, then speeds up by the non-standard $\gamma(1+\beta)$ factor on the $FG$ segment, and slows down again by the standard $\gamma$ factor on the final $GC$ segment. However, we hope it is clear from our discussion that the oddities of the $t^\prime(t)$ function have nothing to do with the actual behavior of either twin's clock --- they are purely coordinate artifacts. One of the oddities of using radar coordinates is that the clock of the twin on Earth "changes" its mode from "slowing down" to "speeding up" when the clock of the twin in the rocket shows $t^\prime=\gamma(1-\beta)T$, that is, this happens {\it before} the twin on the rocket decides at the instant of his/her proper time $t^\prime=\gamma T$ to change direction of the flight \cite{Mcdonald_2020} (and hence experiences acceleration --- another sign that acceleration does not play a significant role in the twin paradox, its only role is to mark a world line that is not a straight line). You can choose other coordinates without this strangeness, but this will lead to other artifacts (see, for example, \cite{Unruh_1981} or \cite{Gron_2006}).

In any case, no matter what coordinates the flying twin chooses, if he/she does everything correctly, the end result will be the same: $t^\prime(2T)=2T/\gamma$, that is, the traveling twin will return younger than his/her earthly twin. This is a simple consequence of the triangle inequality in Minkowski geometry: among timelike worldlines connecting two fixed points in Minkowski spacetime, a straight line (geodesic) corresponds to the largest proper time interval. In the context of the twin paradox, Malament aptly formulates this inequality as follows: "saving time costs money", since acceleration requires fuel, which is not free (note to those who decide to save time in this way: in general spacetime this is only true locally) \cite{Malament_2005}.

\section{Concluding remarks}
``Special Relativity is a very simple theory that is commonly presented in a complex and confusing way" \cite{Maudlin_2012}. Inspired by the legend of the Chukchi shaman, we sought to determine whether a conceptualization of elementary special relativity could be based on a nonintegrable proper time. We identified three key components upon which such a conceptualization could be based: 1) a four-dimensional spacetime as an arena in which physical events can occur; 2) the structure of a light cone at each point in spacetime, which delimits the domain of causal influence associated with a given event in spacetime; 3) and the clock hypothesis, according to which an ideal clock measures a proper time, which represents the "length" of a timelike worldline. In my opinion, none of these basic concepts of special relativity carries any paradoxical connotation for students, except perhaps the non-integrable nature of proper time. However, as the Chukchi myth testifies, Newtonian absolute time is a late scientific and cultural construct, and not our innate perception of time.

Real clocks are complex objects, and the desire to avoid using their idealization as a fundamental element of spacetime theory is understandable. This explains axiomatization schemes that introduce a metric structure to spacetime not through ideal clocks, but rather through worldlines of freely falling particles, defining a projective structure compatible with the conformal structure of light cones (for example, the famous Ehlers-Pirani-Schild axiomatization \cite{Ehlers_1972}). However, the introduction of proper time as an independent basic element of the theory of relativity emphasizes the special role that time plays in the theory: it is not simply a fourth dimension.

\section*{Declaration of competing interest}
The author declares that has no known competing financial interests or personal relationships that could have appeared to influence the work reported in this paper.

\section*{Acknowledgments}
The author would like to express his gratitude to Lyudmila Georgievna Tagiltseva, who introduced him to Tan-Bogoraz and his description of the clock paradox in Chukchi myth. He also thanks to Kirk McDonald for helpful correspondence, and the anonymous reviewers for constructive comments.

\bibliography{clocks.bib}
\end{document}